\documentclass[aps,prc,twocolumn,unsortedaddress,showpacs]{revtex4-1}

\usepackage{graphicx}

\begin{document}


\title{Intermittency route to chaos for the nuclear billiard - a qualitative study}

\author{Daniel Felea}
\email{dfelea@spacescience.ro}
\affiliation{Institute of Space Sciences, P.O.Box MG 23, RO 77125, Bucharest-M\u{a}gurele, Romania}

\author{Ion Valeriu Grossu}
\author{Cristian Constantin Bordeianu}
\author{C\u{a}lin Be\c{s}liu}
\author{Alexandru Jipa}
\affiliation{Faculty of Physics, University of Bucharest, P.O.Box MG 11, RO 77125, Bucharest-M\u{a}gurele, Romania}

\author{Aurelian-Andrei Radu}
\author{Ciprian-Mihai Mitu}
\author{Emil Stan}
\affiliation{Institute of Space Sciences, P.O.Box MG 23, RO 77125, Bucharest-M\u{a}gurele, Romania}

\date{\today}

\begin{abstract}
We analyze on a simple classical billiard system the onset of chaotical behaviour in different dynamical 
states. A classical version of the "nuclear billiard" with a $2D$ deep Woods-Saxon potential is used. 
We take into account the coupling between the single-particle and the collective degrees of freedom 
in the presence of dissipation for several vibrational multipolarities. For the considered oscillation 
modes an increasing divergence of the nucleonic trajectories from the adiabatic to the resonance regime 
was observed. Also, a peculiar case of intermittency is reached in the vicinity of the resonance, for 
the monopole case. We examine the order-to-chaos transition by performing several types of qualitative 
analysis including sensitive dependence on the initial conditions, single-particle phase space maps, 
fractal dimensions of Poincare maps and autocorrelation functions.
\end{abstract}

\pacs{24.60.Lz, 05.45.-a, 05.45.Pq, 21.10.Re}

\maketitle

\section{\label{intro}Introduction}
Deterministic chaos, is usually defined as irregular, unpredictable behaviour of the trajectories 
generated by nonlinear systems whose dynamical laws, involving no randomness or probabilities, 
predict a unique time evolution of a given system.

Over the last two decades an increasing number of papers have treated the study of the deterministic 
chaotical behaviour of Fermi nuclear systems \cite{burgio-95,baldo-96,baldo-98,blocki-78,ring-80,speth-81,%
wong-82,grassberger-83,sieber-89,rapisarda-91,abul-magd-91,blocki-92,blumel-92,baldo-93,blocki-93,berry-93,%
ott-93,bauer-94,hilborn-94,blumel-94,drozdz-94,drozdz-95,bauer1-95,jarzynski-95,bulgac-95,atalmi-96a,atalmi-96b}. 
The interest for analyzing the order-to-chaos transitions on such systems was linked to the problem of 
the onset of dissipation of collective systems through mainly one-body and two-body processes. Among these 
we mention the interaction of the nucleons with the potential well, the evaporation of individual nucleons 
in nuclear peripheral interactions, and the collisions between nucleons without taking into account the 
Pauli blocking effect.

These kinds of analyses were performed for the first time by Burgio, Baldo \textit{et al.} \cite{burgio-95,%
baldo-96,baldo-98} considering a system of nucleons which move within a container modelled as a Woods-Saxon 
type potential and kick the container walls with a specific frequency. They discuss the damping of the 
movement and the relation with order-chaos transition in single-particle dynamics.

On that classical model worked Papachristou and collab. \cite{papachristou-08}, studying the decay width 
of the Isoscalar Giant Monopole Resonance for various spherical nuclei. Following also that formalism, 
the beginning of the chaotic behaviour for a number of nucleons in various dynamical regions at several 
multipolarities was surveyed \cite{felea-01,felea-02,bordeianu-08a,bordeianu-08b,bordeianu-08c}.

In this paper, we investigated the chaotic behaviour of a single-nucleon in a two-dimensional ($2D$) 
deep Woods-Saxon potential well for specific physical phases. A qualitative picture of the achievement 
of deterministic chaos was shown for a comparative study between the adiabatic and the resonance stage 
of the nuclear interaction.

Close to resonance we obtained characteristics of the intermittency regime, \textit{i.e.} sudden change 
to a laminar behaviour (so-called intermission) of a specific signal between two turbulent phases, which 
has been detected over and over again in a plethora of experiments regarding the Rayleigh-Benard convection, 
the driven nonlinear semiconductor oscillator, the Belousov-Zhabotinskii chemical reaction, and the 
Josephson junctions (for e.g., \cite{berge-80,pomeau-81,linsay-81,testa-82,jeffries-82,dubois-83,yeh-83,%
huang-87,richetti-87,kreisberg-91}).

Albeit intermittency is a well-known phenomenon for billiards \cite{dahlqvist-92,dahlqvist-95,artuso-96,%
altmann-07}, in particular for Hamiltonian systems with divided phase space (e.g., mushroom \cite{bunimovich-01,bunimovich-03,altmann-05,altmann-06,porter-06} and annular billiards \cite{saito-82}), 
and for connected Hamiltonian systems \cite{markus-03}, we showed that this property also holds for a 
Woods-Saxon billiard container with inelastic particle-wall interactions.

\section{\label{sec:1}Basic formalism}
We chose for the present analysis a simple dynamical system as considered in several previous papers by 
Burgio and collab. \cite{burgio-95,baldo-96,baldo-98}. This system contains a number of $A$ spinless and 
chargeless nucleons, with no internal structure. The nucleons move in a $2D$ deep Woods-Saxon potential 
well considered as a "nuclear billiard" and hit periodically the oscillating surface of the well with a 
certain frequency. The Bohr Hamiltonian of such a system in polar coordinates is considered as:

\begin{eqnarray}
H\left( r_{j},\theta _{j},\alpha \right) &=&{\sum_{j=1}^{A}}\left[ \frac{p_{r_{j}}^{2}}{2m}+%
\frac{p_{\theta _{j}}^{2}}{2mr_{j}^{2}}+V\left(r_{j},R\left( \theta _{j}\right) \right) \right]  \nonumber \\
&&+\frac{p_{\alpha }^{2}}{2M}+\frac{M\Omega ^{2}\alpha ^{2}}{2},
\end{eqnarray}

where $\left\{ p_{r_{j}},p_{\theta _{j}},p_{\alpha }\right\} $ are the conjugate momenta of the particle and 
collective coordinates $\left\{r_{j},\theta _{j},\alpha \right\} $, the nucleon mass $m$ is $938\ \rm{MeV}$, 
$\Omega $ is the oscillating frequency of the collective variable $\alpha $, and $M=mAR_{0}^{2}$ is the Inglis mass.

The Woods-Saxon potential is constant inside the billiard and a very steeply rising function on the surface:

\begin{equation}
V\left( r_j,R\left( \theta _j\right) \right) = \frac{V_0}{1+\exp \left[ \frac{r_j-R\left( \theta _j,\alpha \right) }a\right] },
\end{equation}

with $V_{0}=-1500\ \rm{MeV}$, deep enough to prevent the escape of the nucleons regarded as classical objects 
for the present analysis. For the same reason, the diffusivity coefficient $a$ has a very small value $0.01\ \rm{fm}$. 
The vibrating surface can be written as in \cite{burgio-95,baldo-96,baldo-98}, depending on the collective 
variable and Legendre polynomials $P_{L}\left( \cos \theta_{j}\right)$:

\begin{equation}
R_j = R\left( \theta _j,\alpha \right) = R_0\left[ 1+\alpha P_L\left( \cos \theta_j\right) \right] ,
\end{equation}

where $R_{0}=6\ \rm{fm}$, and $L$ the multipolarity vibration degree of the potential well is $0$ for the 
monopole, $1$ for the dipole, and $2$ for the quadrupole case.

Once the hamiltonian is chosen, the numerical simulations are based on the solution of the Hamilton equations:

\begin{eqnarray}
\stackrel{\cdot }{r_j} &=&\frac{p_{r_j}}m,\ \stackrel{\cdot }{p_{r_j}} = \frac{p_{\theta _j}^2}{mr_j^3}-\frac{\partial V}{\partial r_j}, \\
\stackrel{\cdot }{\theta _j} &=&\frac{p_{\theta _j}}{mr_j^2},\ \stackrel{\cdot }{p_{\theta _j}} = -\frac{\partial V}{\partial R_j} \cdot \frac{\partial R_j}{\partial \theta _j}, \\
\stackrel{\cdot }{\alpha } &=&\frac{p_\alpha }M,\ \stackrel{\cdot }{p_\alpha} = -M\Omega ^2\alpha - \sum_{j=1}^{A}\left(\frac{\partial V}{\partial R_j} \cdot \frac{\partial R_j}{\partial \alpha }\right).
\end{eqnarray}

The Hamilton equations were solved with a Runge-Kutta type algorithm (order 2-3) having an optimized step 
size and taking into account that the absolute error for each variable is less than $10^{-6}$. Total energy 
was verified to be conserved with high accuracy to a relative error level of $10^{-8}$.

The equilibrium deformation parameter $\stackrel{-}{\alpha }$, which is the mean collective variable, can 
be calculated (for e.g., for $L = 0$) by equating the mechanical pressure of the wall, $P_{wall}$:

\begin{equation}
P_{wall}=\frac{M\Omega ^{2}}{2\pi R_{0}^{2}}\cdot \frac{\stackrel{-}{\alpha }}{1+\stackrel{-}{\alpha }}.
\end{equation}

The pressure exerted by the particles is $P_{part}$, $\rho $ denotes the particle density, and $T$ is the 
apparent temperature of the system that equals the 2D kinetic energy, using the natural system of units 
($\hbar=c=k_{B}=1$):

\begin{equation}
P_{part}=\rho T=\frac{AT}{\pi R_{0}^{2}\left( 1+\stackrel{-}{\alpha }\right) ^{2}}.
\end{equation}

Thus, one gets the equation for the equilibrium value of the collective coordinate in 2D, in the monopole case:

\begin{equation}
\stackrel{-}{\alpha }\left( 1+\stackrel{-}{\alpha }\right) =\frac{2T}{mR_0^2\Omega ^2}.
\end{equation}

Then a small perturbation of this collective variable was considered $\alpha \left( t=0\ \rm{fm}/c\right) =\ %
\stackrel{-}{\alpha }+0.15$ \cite{burgio-95,baldo-96,baldo-98} and the evolution of the physical system was 
thoroughly investigated.

\section{\label{sec:2}The qualitative analysis of the route to chaos}

\subsection{On the resonance condition}

One can choose the wall oscillation taking place close to adiabatic conditions, imposing a wall frequency 
smaller than the single-particle one. Thus, the frequency of vibration $\Omega _{ad}$ was chosen less 
than $0.05\ c/\rm{fm}$, which corresponds to an oscillation period equal to:

\begin{equation}
\tau_{wall}=\frac{2\pi }{\Omega _{ad}}\!\geq\!125.66\ \rm{fm}/c.
\end{equation}

By introducing the maximum particle speed:

\begin{equation}
v=\sqrt{\frac{2T}m},
\end{equation}

and the parameters as in \cite{burgio-95,baldo-96,baldo-98}: $R_{0}=6\ \rm{fm}$ and $T=36\ \rm{MeV}$, one 
can obtain the value for the single-particle period:

\begin{equation}
\tau_{part}=\frac{2R_0}v\approx 43.33\ \rm{fm}/c.
\end{equation}

In addition to \cite{burgio-95,baldo-96,baldo-98} we introduced a physical constraint to this elementary 
physical system and continued that type of analysis necessary for the study of a nonintegrable dynamical 
system. At the beginning we considered a physical situation and we chose instead a static vibrating 
"nuclear billiard", a projectile nucleus having the same properties, colliding with a target nucleus. 
It is well-known that the nuclear interaction, at incident energies ranging from \rm{MeV} to \rm{GeV}, 
can result in a multitude of processes from the nuclear evaporation to complete fragmentation or 
multifragmentation, according to the impact parameter.

It was shown in \cite{bauer2-95,felea-99} that during this kind of processes even for peripheral events 
an unnegligeable amount of energy is transferred by nucleon-nucleon scattering to the nucleons of the 
projectile and not only the transverse momentum distributions, but also the longitudinal momentum 
distributions as measured in the projectile fragmentation rest frame can reveal the centrality status 
of the interaction. It can also offer a hint on the apparent temperature of a Fermi gas of nucleons 
which was found to be \cite{felea-99} near the isotopic temperatures, \textit{i.e.} several \rm{MeV} 
\cite{pochodzalla-87,kunde-91,morrissey-94,serfling-98}.

It was therefore supposed that the target fragmentation can be associated with a resonance process. In 
order to obtain such behaviour, the wall frequency was gradually increased to the resonance frequency 
$\Omega _{res}=0.145\ c/\rm{fm}$. However, nuclear evaporation or plain breakup of a projectile nucleus 
can take place long before this regime is achieved by redistributing energy between the nucleons themselves 
and also between single-particle degrees of freedom and collective ones. Individual nucleons or clusters 
can thus have enough kinetic energy to escalade the wall barrier.

We should also emphasize that we can either have the case that can be put in correspondence with a 
nuclear collision process, \textit{i.e.} the variation of the nucleonic frequency oscillation as the 
apparent temperature of the nucleons in the nuclei increases (Eqs. 11 and 12), maintaining the potential 
well vibration constant, or respectively, the inverse situation in which the period between two consecutive 
collisions of the nucleon with the self-consistent mean field is kept invariable, while modifying the 
oscillation modes of the nuclear surface. The latter regards our studied case and is the reversed physical 
case previously described. It was used because of the specific choice of the "toy model" parameters 
described in \cite{burgio-95,baldo-96,baldo-98}.

But the most realistic evolution of the nucleons in a chosen potential can assume a simultaneous variation 
of both angular frequencies. The resonance condition of the coupled classical oscillators should remain 
however an important condition for a rapid appearance of a deterministic chaotical behaviour of the 
physical system in study at different time scales. A proper analysis of a system should provide the 
variation of the collision radian frequency of the nucleons inside the "billiard" as the apparent 
temperature increases and the change in the vibrating potential period, supposing that the multipolarity 
increases when pumping energy in the "nuclear reservoir" during interaction. We can for example use in 
simulations, for nuclei with a large number of nucleons, the Liquid Drop Model or the Collective Model, 
which predict a frequency of vibration as function of the multipolarity deformation degree:

\begin{equation}
\Omega _{L}=\sqrt{\frac{C_{L}}{B_{L}}},
\end{equation}

with $C_{L}$ being the elasticity coefficient, and $B_{L}$ the mass coefficient for the oscillator of 
$L$ multipolarity.

We will briefly discuss on how the resonance condition might look like and for which particular case(s) 
it can be applied. By reverting to the set of differential nonlinear Hamilton equations and combining 
the last two of them (see Eq. 6), it emerged:

\begin{equation}
\stackrel{..}{\alpha }+\Omega^{2}\alpha = -\frac{V_0 R_0}{a M} \cdot \sum_{j=1}^{A} P_L\left( \cos \theta_j\right) \frac{e^{\frac{r_j-R_j}{a}}}{\left[1+e^{\frac{r_j-R_j}{a}}\right]^{2}}.
\end{equation}

One can now compare the resulted equation with a harmonic oscillator system in the presence of dissipative 
processes ("damped oscillator") and of a external harmonic driving force of $F_0$ amplitude:

\begin{equation}
\stackrel{..}{\alpha }+\Gamma \stackrel{.}{\alpha } +\Omega^{2}\alpha = -\frac{F_0 R_0}{M} \cdot cos\: \omega t.
\end{equation}

In a first approximation we can neglect the damping constant $\Gamma$, as noticed from the Figs. \ref{fig:1}-\ref{fig:4}. 
Besides, dissipation will appear if only the collective coordinate $\alpha$ is averaged over a large 
number of events, for all multipolarities considered \cite{burgio-95,baldo-98}. For the uncoupled 
equations ($UCE$) case there is no damping either, as the solution of a homogeneous Eq. 14 is a purely 
harmonic one: $\alpha = \alpha_{0}\: cos\: \Omega t$.

Also, we remark that for the dipole collective oscillations: $P_1\left( \cos \theta_j\right)=cos\: %
\omega_{j} t$ and that, for a statistical ensemble of $A$ nucleons, the one-particle radian frequency 
can be approximated with a constant, $\omega_j \approx \omega$. For the Woods-Saxon potential energy: 
$V(t)=\sum_{j=1}^{A} V(r_j,\theta_j,\alpha)$, it can always be derived a conservative force $F_0$:

\begin{equation}
F_0\left(t\right) = -\sum_{j=1}^{A} \frac{\partial V\left(t\right)}{\partial r_j} = \frac{V_0}{a} \cdot \sum_{j=1}^{A} \frac{e^{\frac{r_j-R_j}{a}}}{\left[1+e^{\frac{r_j-R_j}{a}}\right]^{2}}.
\end{equation}

A similarity between Eq. 14 and 15 was eventually obtained, because once shifting towards collectives 
of nucleons, the total force becomes a sum over a number of time dependent forces, each acting on a 
single nucleon. Therefore, the net force for a pack of nucleons tends to vary little with time: $F_0 %
\neq F_0(t)$, and the tendency increases with $A$.

In conclusion, the resonance condition, as stated in Eq. 14, can be definitely used for the multi-particle 
couplings with a dipolar shape of the potential well. This was verified by studying the variance of 
the largest Lyapunov exponent and of the Kolmogorov-Sinai entropy with the radian frequency of wall 
oscillation on a ten-nucleons system \cite{bordeianu-08c}. For the other multipole degrees taken into 
account, and as well, for the dipole oscillations coupled with a single-nucleon dynamics, the nonlinear 
character of the differential Equation 14 offers a somewhat intricate perspective over obtaining an 
analytical generalized resonance condition, and for the moment is left aside for further investigation.

As we confine the following analysis on the single-particle chaotical dynamics, the frequency matching 
of the coupled oscillators: $\omega _{part}=\Omega _{res}=0.145\ c/\rm{fm}$, will be generically designated 
as the resonance stage of the interaction.

\subsection{Dependence on the initial conditions}

\begin{figure*}
\resizebox{1.\hsize}{!}{\includegraphics{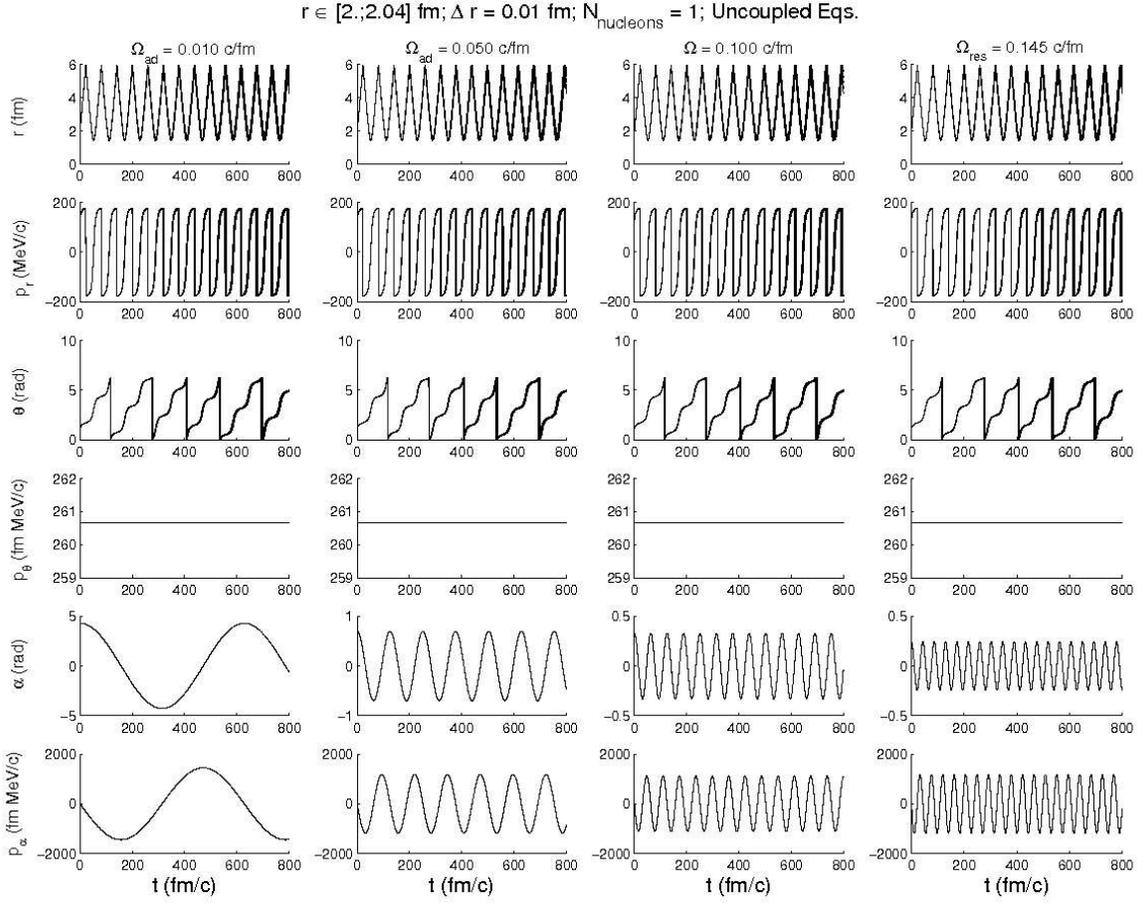} }
\caption{\label{fig:1}The sensitive dependence on the initial small perturbation of the radius parameter (0.01 \rm{fm}/c) when adiabatic and resonance conditions are imposed (uncoupled one-nucleon and collective degrees of freedom case).}
\end{figure*}

\begin{figure*}
\resizebox{1.\hsize}{!}{\includegraphics{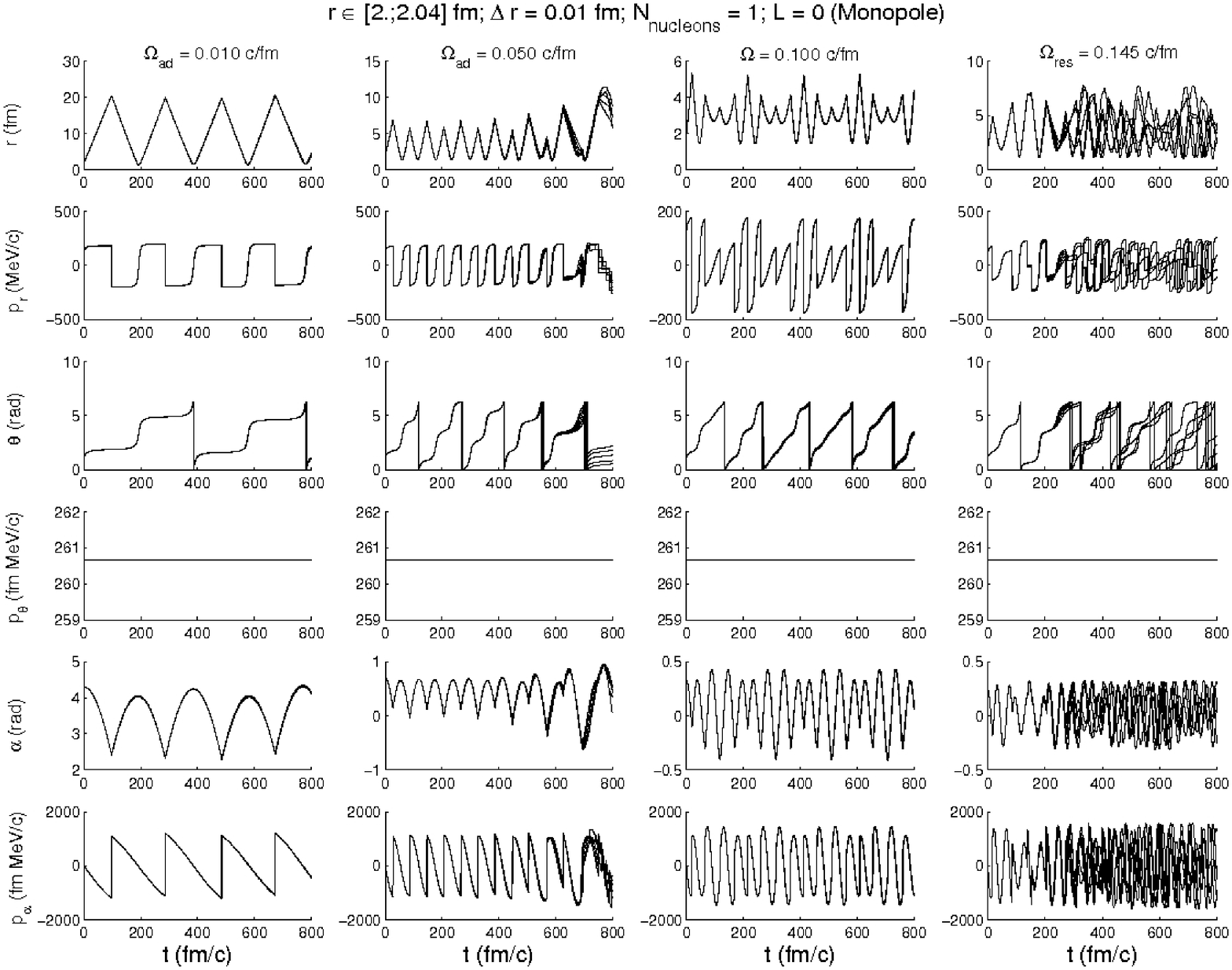} }
\caption{\label{fig:2}The sensitive dependence on the initial small perturbation of the radius parameter (0.01 \rm{fm}/c) when adiabatic and resonance conditions are imposed (monopole case).}
\end{figure*}

\begin{figure*}
\resizebox{1.\hsize}{!}{\includegraphics{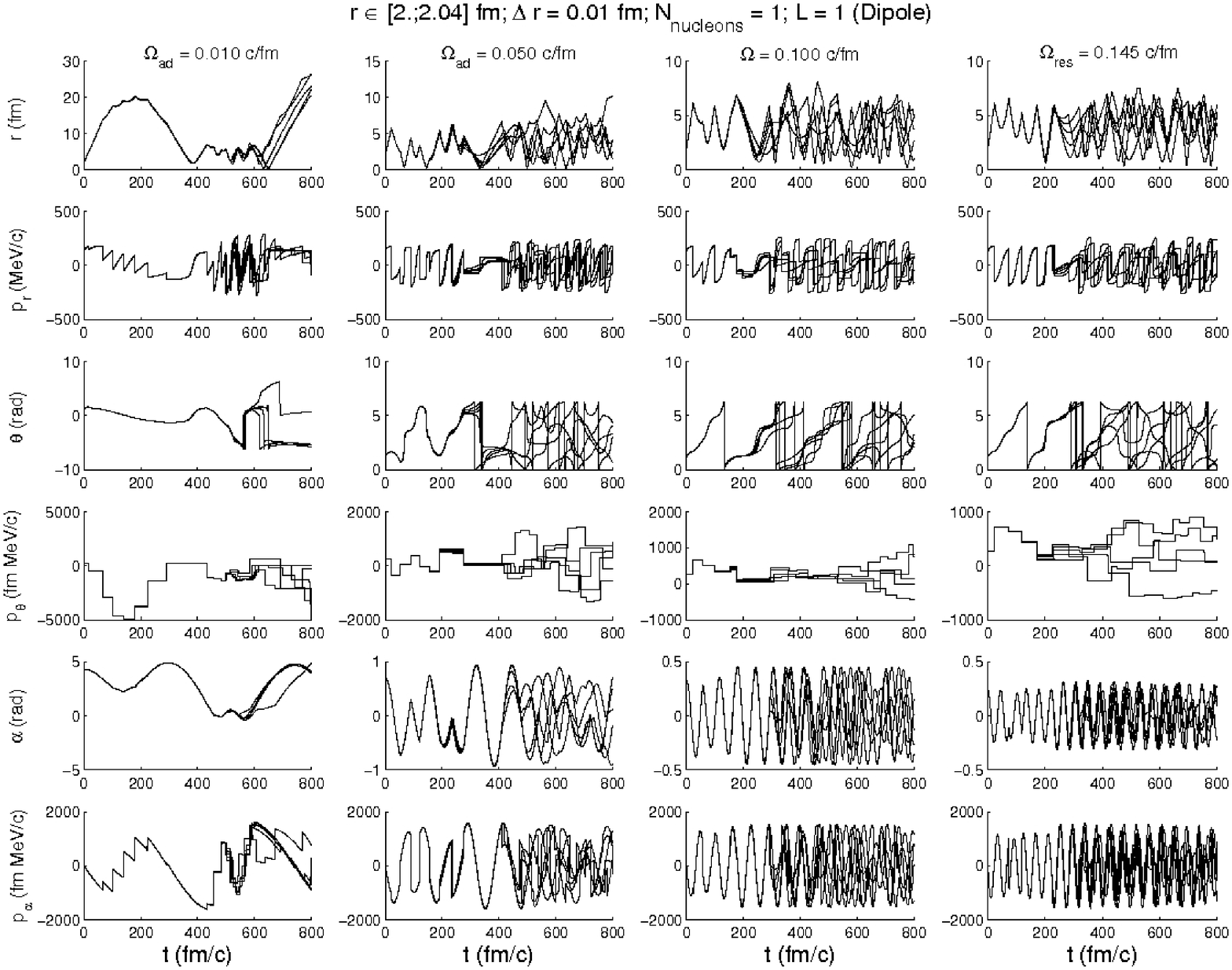} }
\caption{\label{fig:3}The sensitive dependence on the initial small perturbation of the radius parameter (0.01 \rm{fm}/c) when adiabatic and resonance conditions are imposed (dipole case).}
\end{figure*}

\begin{figure*}
\resizebox{1.\hsize}{!}{\includegraphics{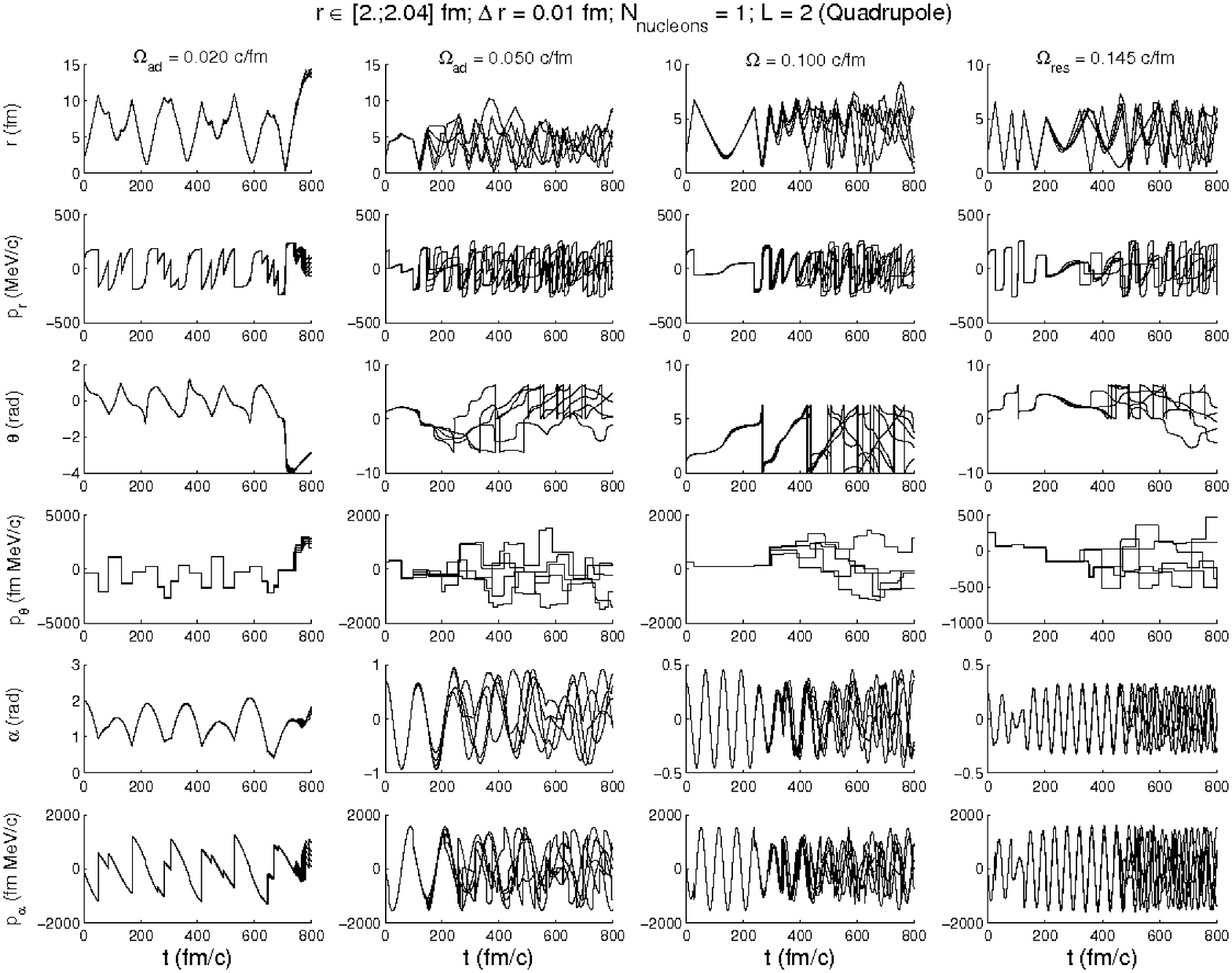} }
\caption{\label{fig:4}The sensitive dependence on the initial small perturbation of the radius parameter (0.01 \rm{fm}/c) when adiabatic and resonance conditions are imposed (quadrupole case).}
\end{figure*}

In order to detect chaos in simple systems, several customary methods are used \cite{schuster-84}. The 
sensitive dependence on the initial conditions, the so-called "Butterfly effect" is the first type of 
analysis presented in this article. For a given multipolarity degree we studied the time dependence of 
both, single-particle and collective variables, with small perturbations applied to a single parameter 
(for e.g., $\Delta r=0.01\ \rm{fm}$). This type of analysis is presented for four cases, ranging from 
$UCE$ to quadrupole deformations of the two-dimensional wall surface, and for the two physical regimes 
in study, adiabatic and of resonance, respectively (Figs. \ref{fig:1}-\ref{fig:4}).

At a first glance, the time dependence of the collective degrees of freedom "looks chaotic" for all 
considered situations with the exception of $\left( 4+2\right) $ uncoupled nonlinear differential 
equations case (Fig. \ref{fig:1}), as one would naturally expect. In this case, a periodical structure 
would definitely appear in sharp contrast with the quasi-periodicity shown for different multipole 
collective modes. Also, the time evolution of the single-particle variables is clearly chaotic in all 
cases where the strong coupling between the $\left( 4+2\right) $ nonlinear differential equations 
describing the single-particle dynamics appears (Figs. \ref{fig:2}-\ref{fig:4}).

The sensitive dependence on the initial conditions, \textit{i.e.} the decoupling of the trajectories at 
macroscopic scale, is found to give the first hints on the behaviour of the nuclear system, in its 
evolution from the quasi-stable states, which can hardly develop a chaotic motion in time (adiabatic 
state), to the unstable ones, characterized by a rapid divergence of the particle trajectories in the 
phase space (resonance regime). Also, in the monopole case the order-to-chaos transition clearly shows 
an intermittent pattern at $\Omega=0.1\ c/\rm{fm}$ (Fig. \ref{fig:2}).

\subsection{Maps of phase space}

Another type of qualitative analysis, indicating different ways toward a chaotic behaviour is based on 
the one-dimensional maps of phase space.

\begin{figure}
\resizebox{1.\hsize}{!}{\includegraphics{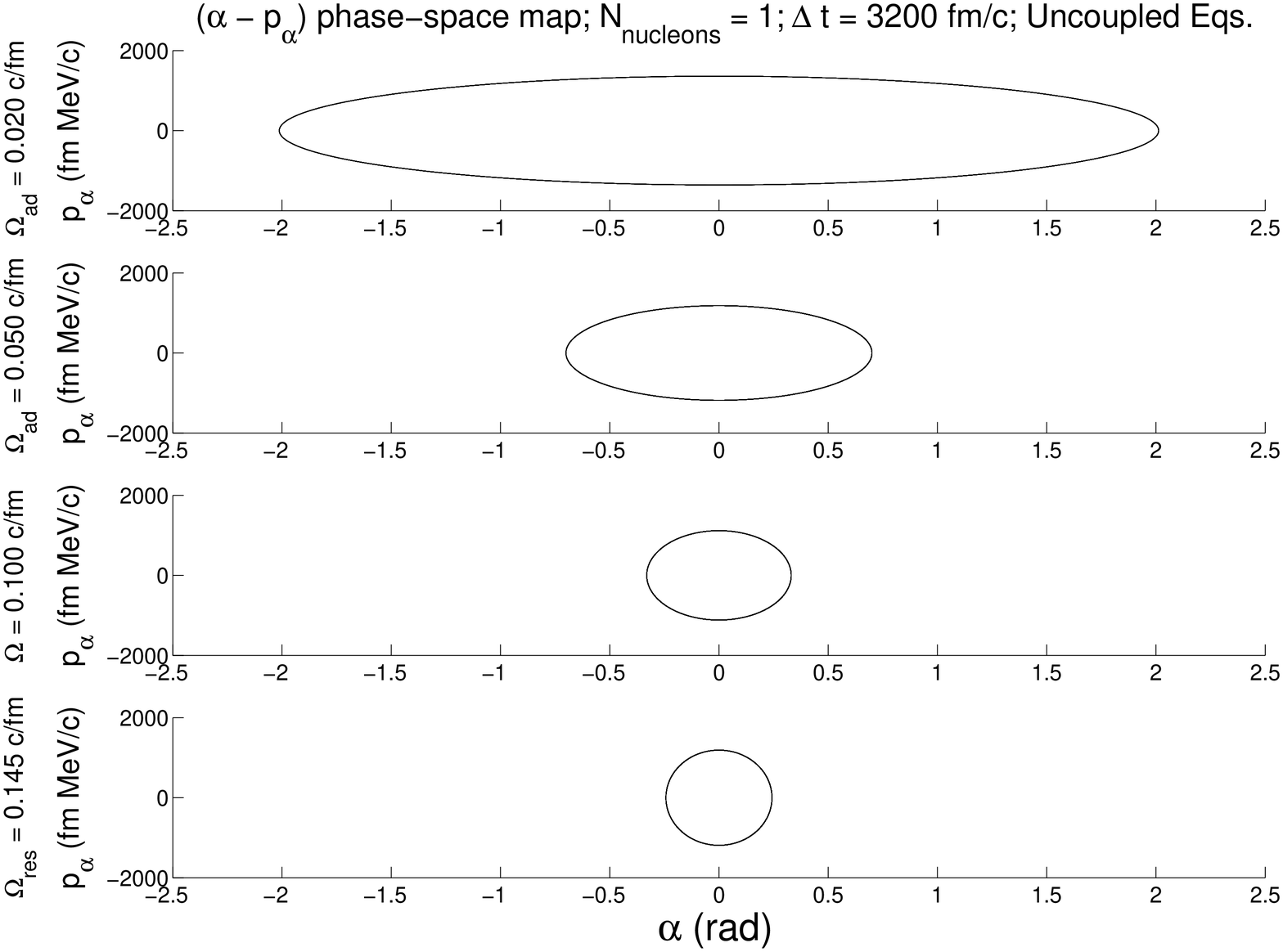} }
\caption{\label{fig:5}The phase space of the collective degrees of freedom ($\alpha - p_{\alpha}$) for different wall frequencies (uncoupled single and collective degrees of freedom case).}
\end{figure}

\begin{figure}
\resizebox{1.\hsize}{!}{\includegraphics{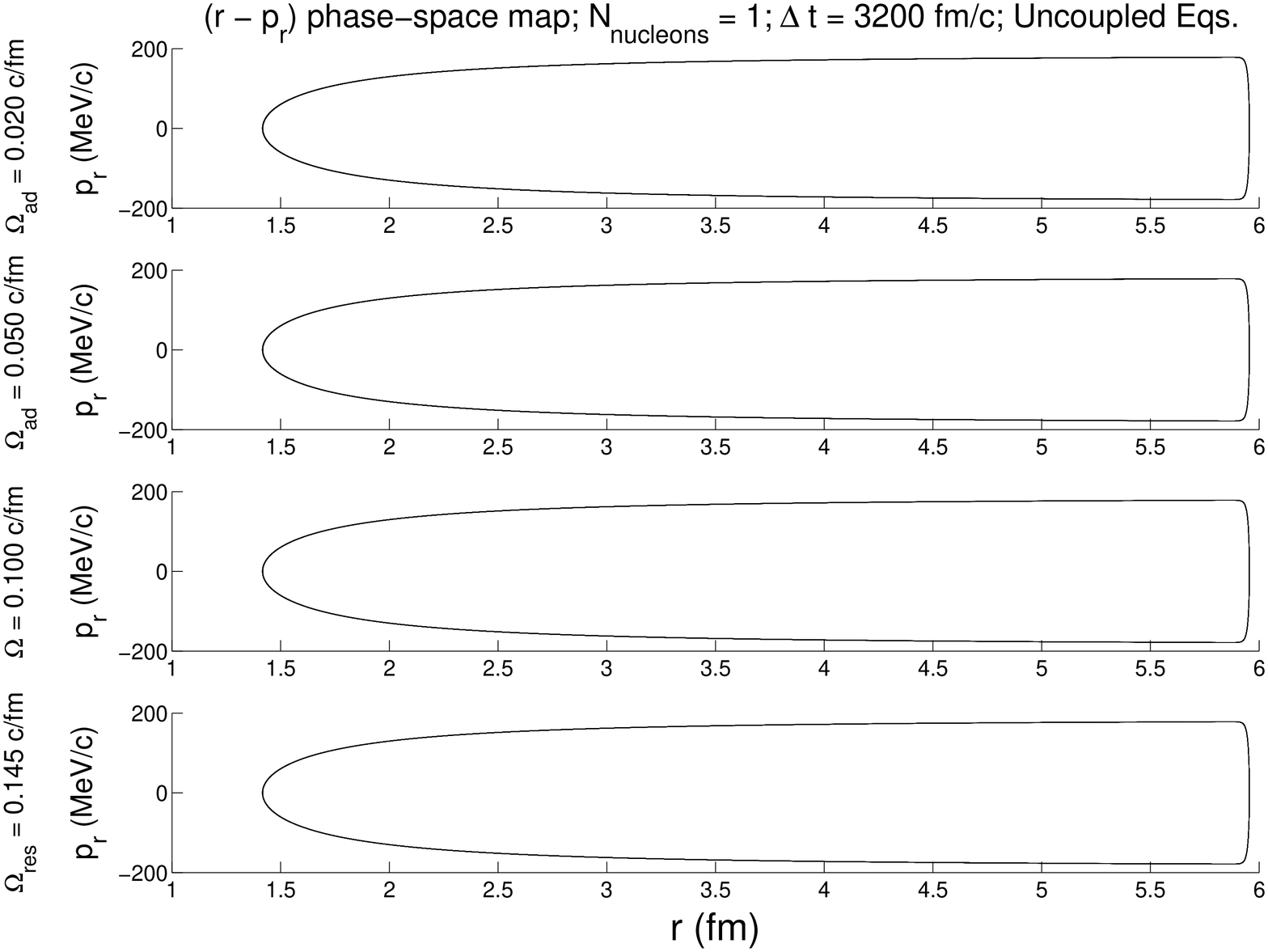} }
\caption{\label{fig:6}The phase space of the single-particle degrees of freedom ($r - p_{r}$) for different wall frequencies (uncoupled single and 
collective degrees of freedom case).}
\end{figure}

\begin{figure}
\resizebox{1.\hsize}{!}{\includegraphics{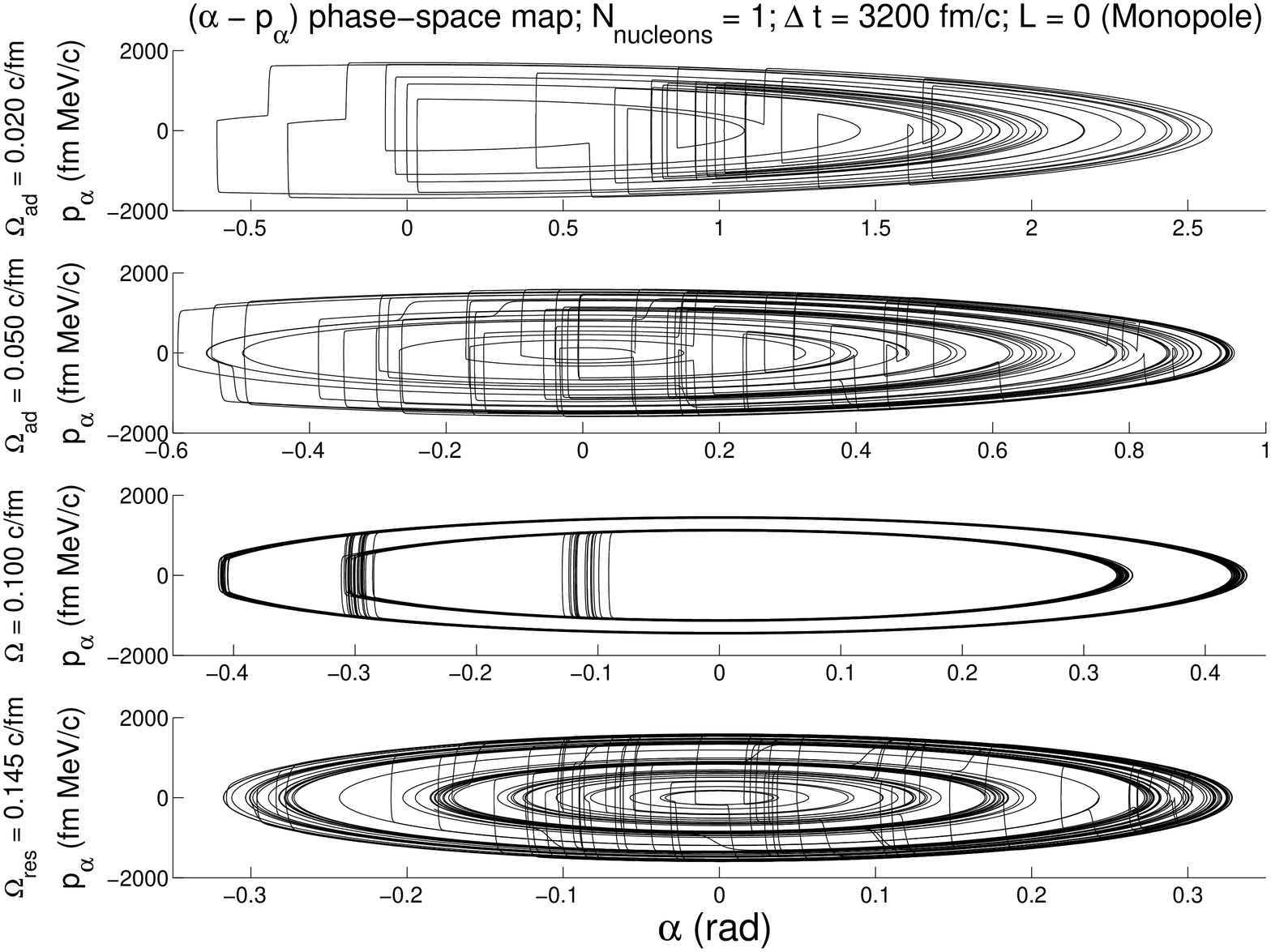} }
\caption{\label{fig:7}The phase space of the collective degrees of freedom ($\alpha - p_{\alpha}$) for different wall frequencies (L = 0).}
\end{figure}

\begin{figure}
\resizebox{1.\hsize}{!}{\includegraphics{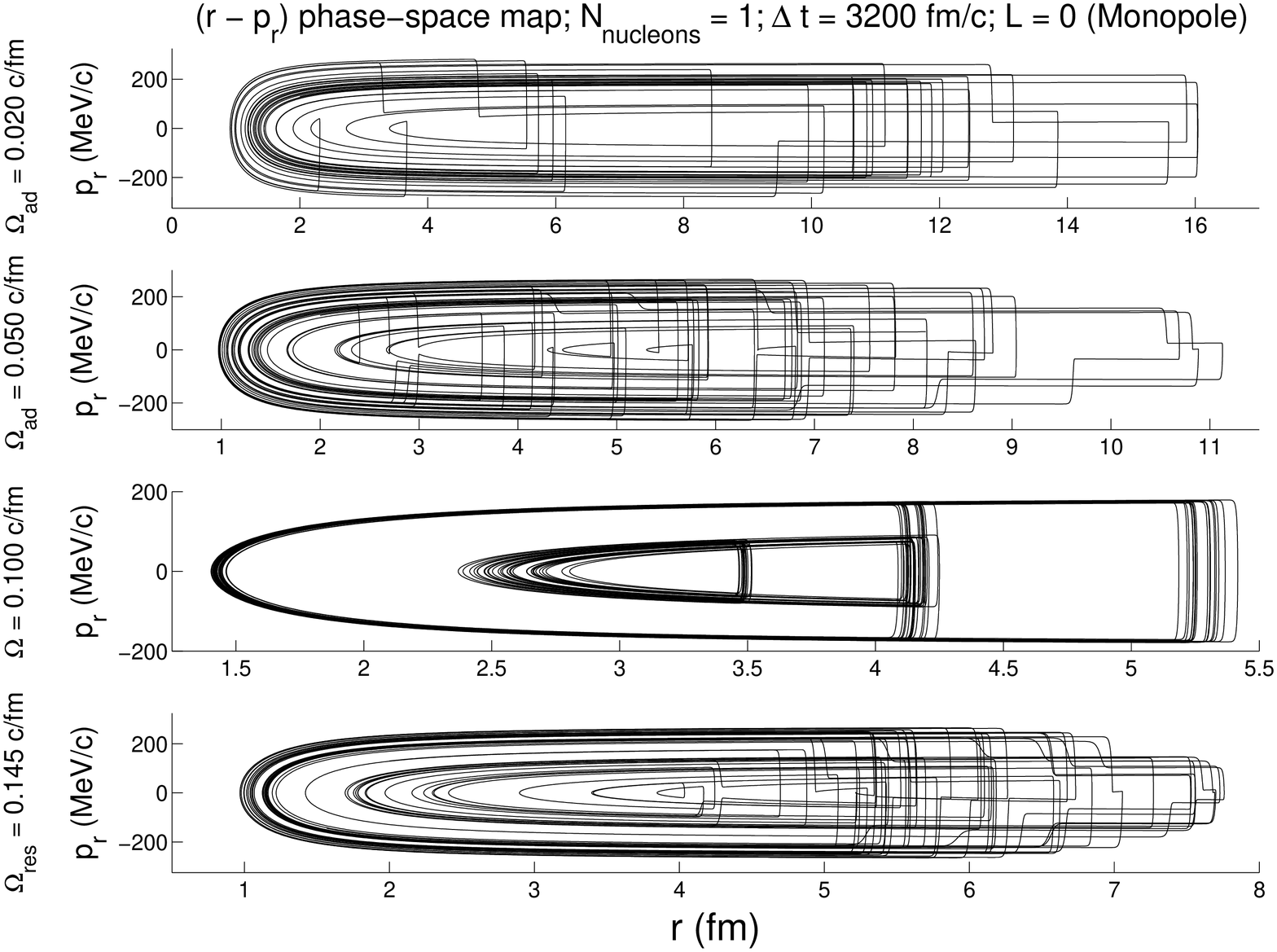} }
\caption{\label{fig:8}The phase space of the single-particle degrees of freedom ($r - p_{r}$) for different wall frequencies (L = 0).}
\end{figure}

\begin{figure}
\resizebox{1.\hsize}{!}{\includegraphics{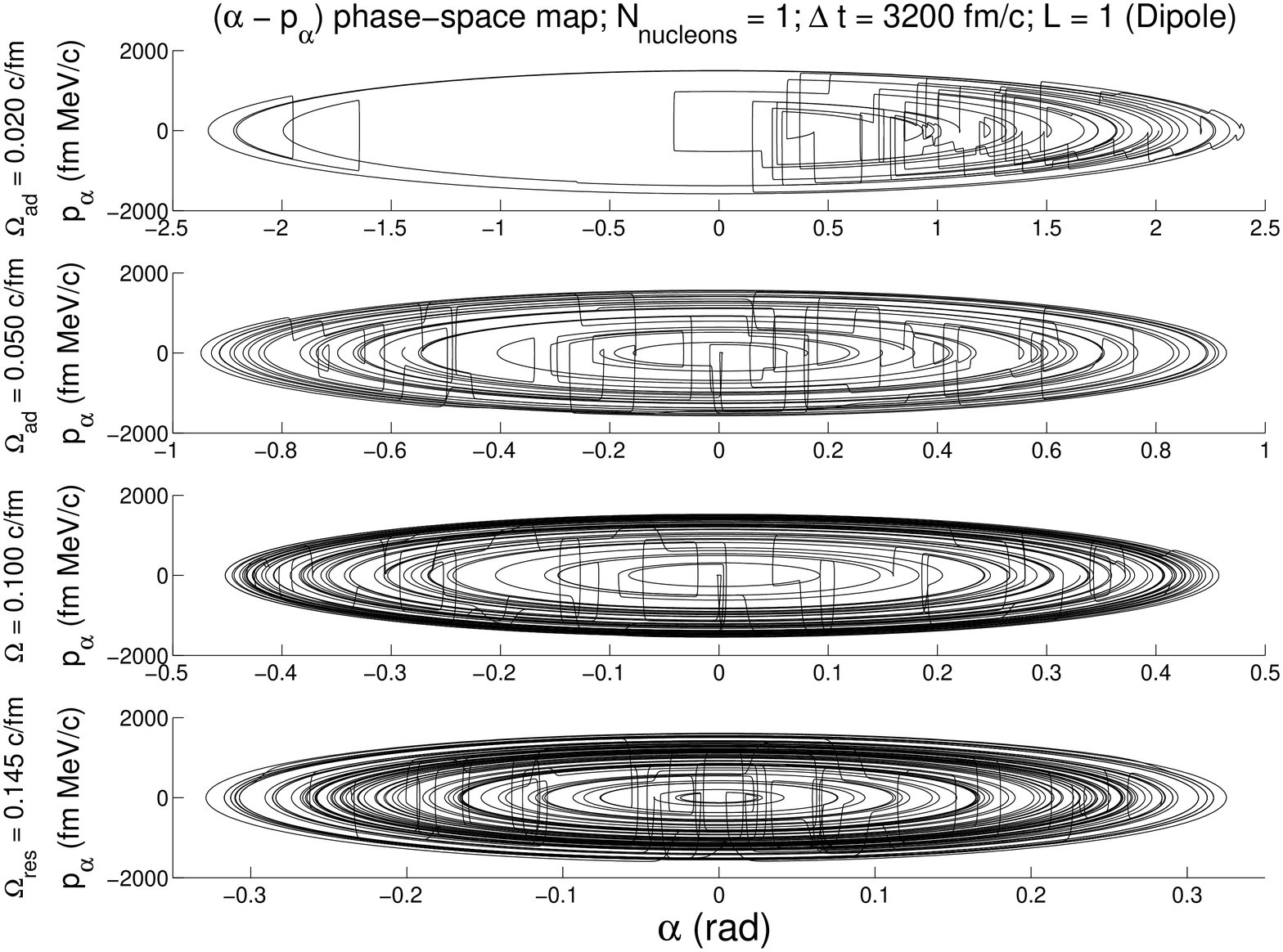} }
\caption{\label{fig:9}The phase space of the collective degrees of freedom ($\alpha - p_{\alpha}$) for different wall frequencies (L = 1).}
\end{figure}

\begin{figure}
\resizebox{1.\hsize}{!}{\includegraphics{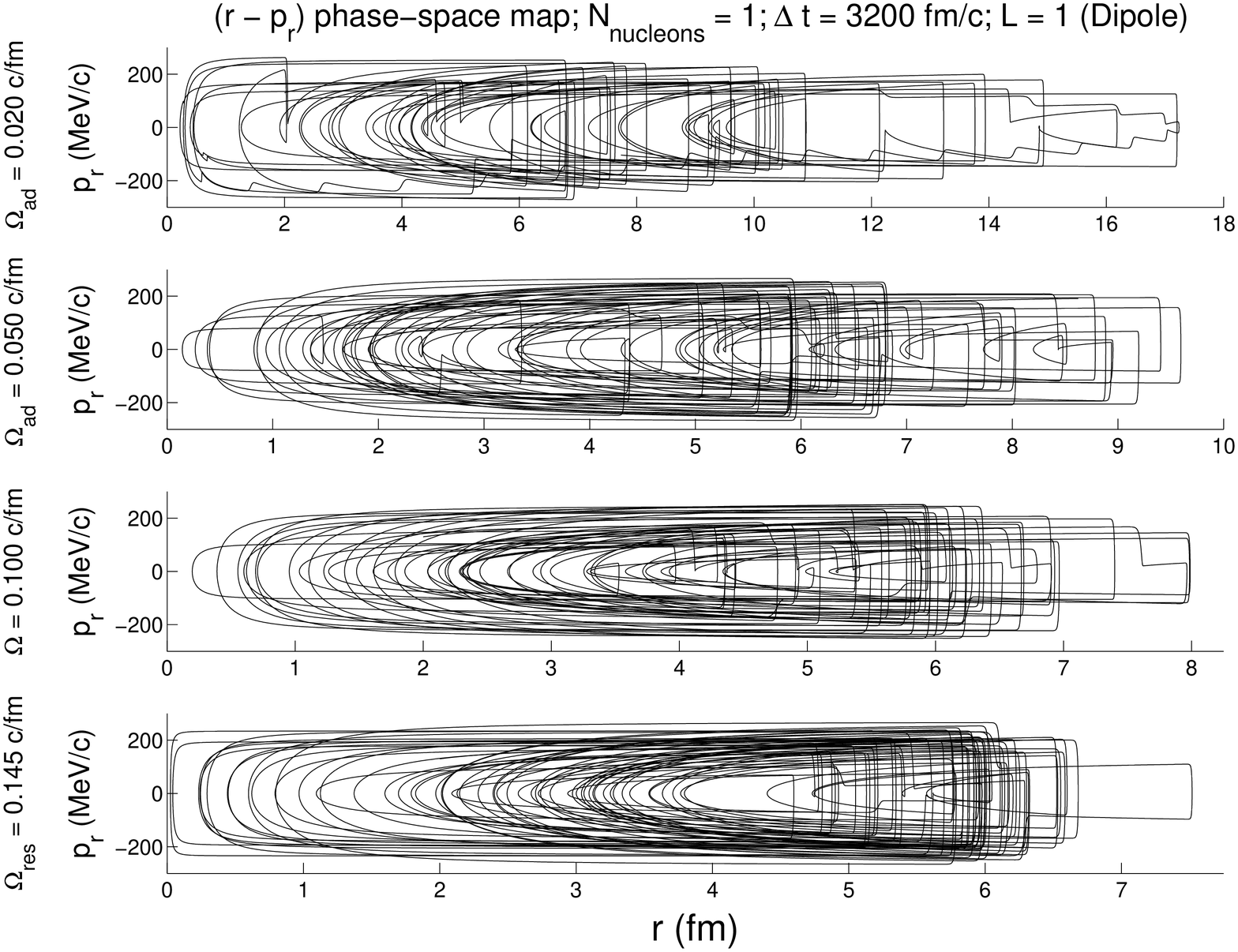} }
\caption{\label{fig:10}The phase space of the single-particle degrees of freedom ($r - p_{r}$) for different wall frequencies (L = 1).}
\end{figure}

\begin{figure}
\resizebox{1.\hsize}{!}{\includegraphics{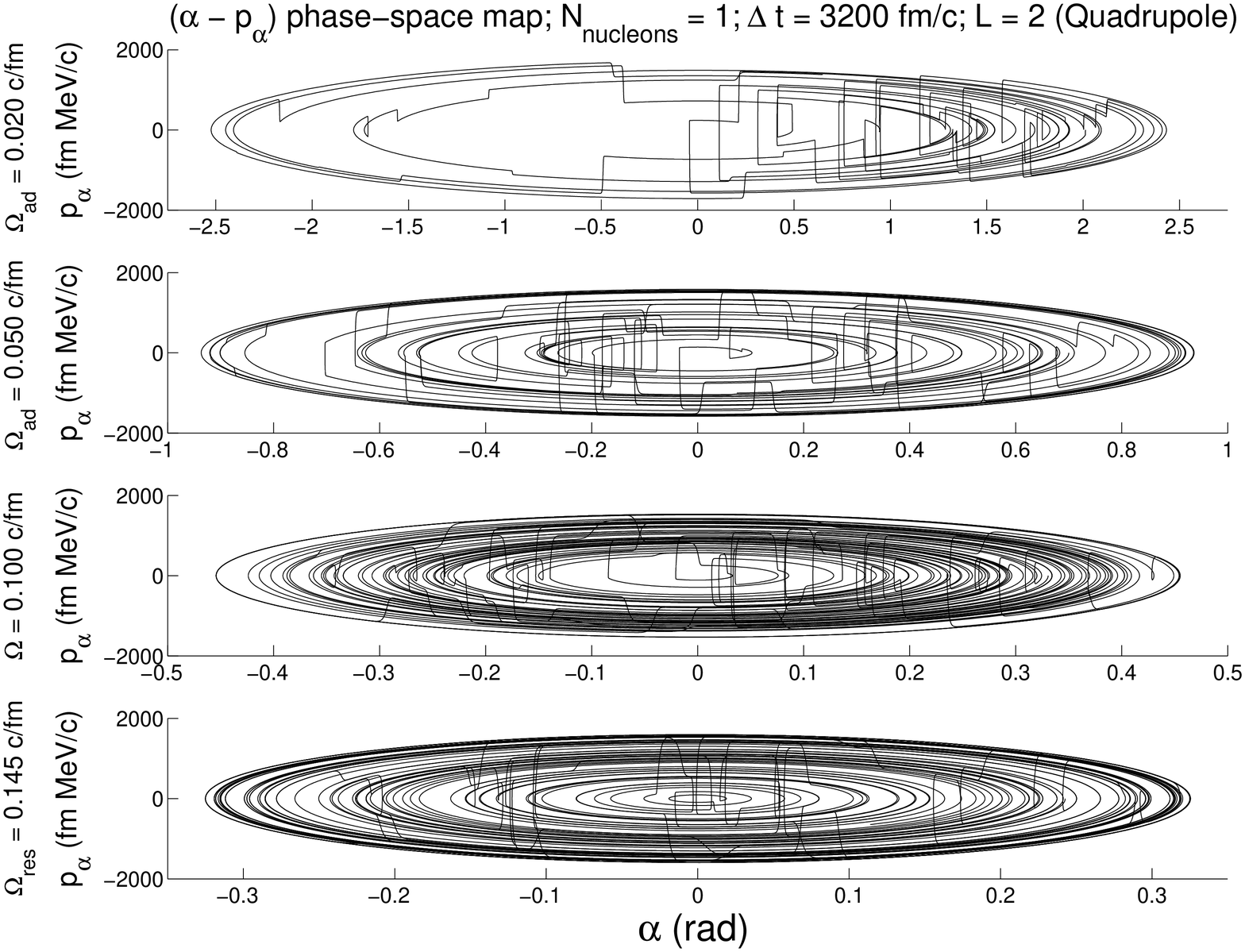} }
\caption{\label{fig:11}The phase space of the collective degrees of freedom ($\alpha - p_{\alpha}$) for different wall frequencies (L = 2).}
\end{figure}

\begin{figure}
\resizebox{1.\hsize}{!}{\includegraphics{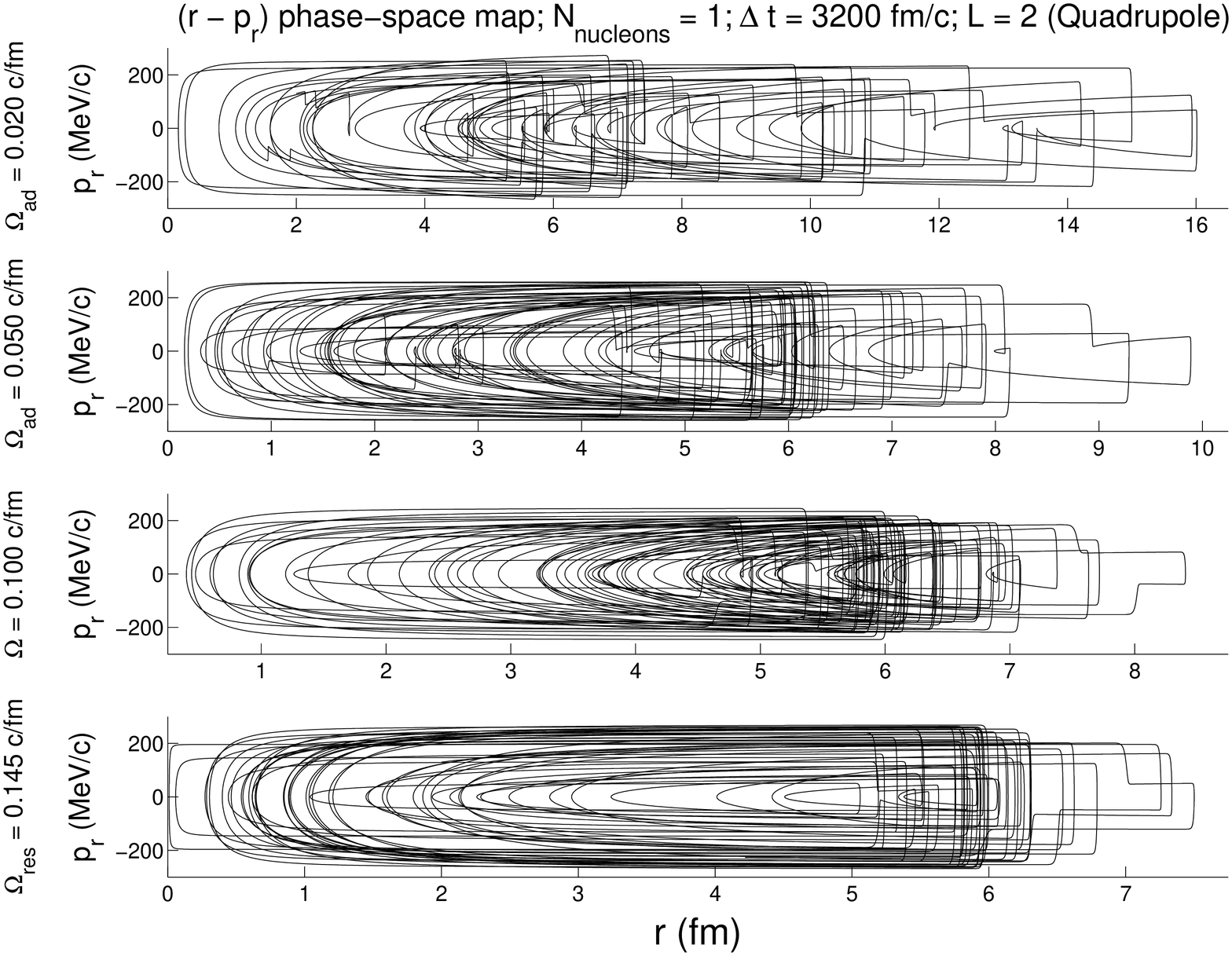} }
\caption{\label{fig:12}The phase space of the single-particle degrees of freedom ($r - p_{r}$) for different wall frequencies (L = 2).}
\end{figure}

Thus, for a temporal scale of $3,200\ \rm{fm}/c$ we represented the evolution of the multipolarity 
deformation degree of the potential well in the phase space of the vibrational variables $\left(\alpha %
\leftrightarrow p_{\alpha }\right)$, as well as in the $1D$ phase space of a single-particle $\left( r%
\leftrightarrow p_{r}\right) $. This type of analysis was performed for the $UCE$, monopole, dipole and 
quadrupole cases, using the same physical conditions, from the adiabatic to the resonance phase (Figs. 
\ref{fig:5}-\ref{fig:12}).

A remark that can be formulated from the study of these phase space maps is related to the form of the 
trajectories specific to those described by a harmonic oscillator, especially for the $\left( \alpha %
\leftrightarrow p_{\alpha }\right) $ maps. It can be easily put to the test (Figs. \ref{fig:5}, \ref{fig:7}, 
\ref{fig:9}, and \ref{fig:11}) that the ellipse area $S_{h\!o}$ is proportionally inverse with the wall 
vibration frequency:

\begin{equation}
S_{h\!o}=\frac{2\pi\cdot E_{coll}}{\Omega},
\end{equation}

where $E_{coll}$ is the sum of the last two terms from $(1)$, being the energy of the collective nucleonic 
motion.

The phase space filling degree raises as the vibration wall frequency moves towards the resonance value 
for all multipolarities took into account (Figs. \ref{fig:7}-\ref{fig:12}), with the exception of the 
uncoupled nonlinear differential equations case (Figs. \ref{fig:5} and \ref{fig:6}). This confirms that 
the characteristic time of the macroscopic decoupling of the nucleon trajectories in the Woods-Saxon 
potential evolves towards smaller values, once $\Omega$ is increased and also that the coupling between 
the collective variable motion and the particle dynamics is essential in amplifying the chaotic behaviour.

An intriguing aspect is revealed by the comparison of the filling degrees of the phase space as a 
function of the nucleon oscillation frequency in the chosen potential. One would expect that the 
trajectories degenerate from a simple orbit towards a compact filling of the 
$\left( \alpha \leftrightarrow p_{\alpha }\right) $ or $\left( r\leftrightarrow p_{r}\right) $ plane. 
Moreover, we observe several attraction basins around a few standard orbits, preeminently in the 
monopole case at $\Omega=0.1\ c/\rm{fm}$, characteristic feature of an intermittent behaviour (Figs. 
\ref{fig:7} and \ref{fig:8}).

\subsection{Fractal dimensions of Poincare maps}

The one-dimensional phase space maps offer a hint on their filling degree in time. In the same manner 
one can use the Poincare maps for the "nuclear billiard" regarded as a deterministic dynamical system. 
When analyzing the behaviour of close trajectories in the phase space, starting from a periodical orbit, 
Poincare showed \cite{poincare-890,poincare-892} that there are only three distinct possibilities: 
the curve can be either a closed one (with a fix distance to a periodical orbit), or it can be a spiral 
that asymptotically wraps/unfolds to/from the periodical trajectory.

In order to simplify the temporal analysis for peculiar orbits, Poincare suggested a simple and effective 
method. By choosing a transversal section with $\left( N-1\right) D$ that intersects a $ND$ geometrical 
variety, the sequence of the intersection points is easier to be studied than the whole $ND$ curve. 
Therefore, the cases previously described for the spiral generate a series of points that draw near in 
or move away from the intersection point of the periodical orbit with the transversal section.

Applying the Poincare maps type analysis to the simple chosen physical system, one would expect that 
the compact filling up of a $1D$ map of phase space for vibration frequencies of the Woods-Saxon well 
that match the one-particle oscillation frequencies inside the "billiard" (Figs. \ref{fig:5}-\ref{fig:12}), 
to reverberate by a large density of chaotic distributed intersection points.

In order to estimate the degree of fractality of such densities, we computed the fractal dimensions $d_{f}$ 
of the Poincare maps with one-nucleon radial degrees of freedom $\left( r\leftrightarrow p_{r}\right)$, 
when choosing for the transversal section the polar pair $\left( \theta\leftrightarrow p_{\theta }\right)$ 
as a constant. As only for the monopole and $UCE$ cases the orbital kinetic momentum is a constant of 
motion, we plotted the one-particle radius as a function of radial momentum when the polar coordinate is 
not kept constant, but quasi-constant in order to increase the probability of intersecting the section 
for a given time evolution of the system:

\begin{equation}
\theta=\theta_{0}\pm\epsilon_{\theta}.
\end{equation}

We then chose a very small value for the $\left( \theta\leftrightarrow p_{\theta }\right)$ thickness: 
$\epsilon_{\theta}=10^{-4}$ radians and let the system evolve for a period of $\Delta t=5\cdot10^{6}\ \rm{fm}/c$.

The values were calculated using the box counting algorithm in a specific Visual Basic 6 application 
\cite{grossu-09a,grossu-09b} on Poincare maps. The fractal dimension of a map covered by $N$ boxes of 
length $r$ is described by the following expression (see, for e.g., \cite{landau-07,landau-08}):

\begin{equation}
d_{f}=\lim_{r\rightarrow 0}\frac{\log N\left(r\right)}{\log \left(1/r\right)},
\end{equation}

where we took: $r=1/2^{n}$, $0\leq n\leq n_{max}$. The maximum value $n_{max}$ denotes the highest 
chosen resolution.

\begin{figure}
\resizebox{1.\hsize}{!}{\includegraphics{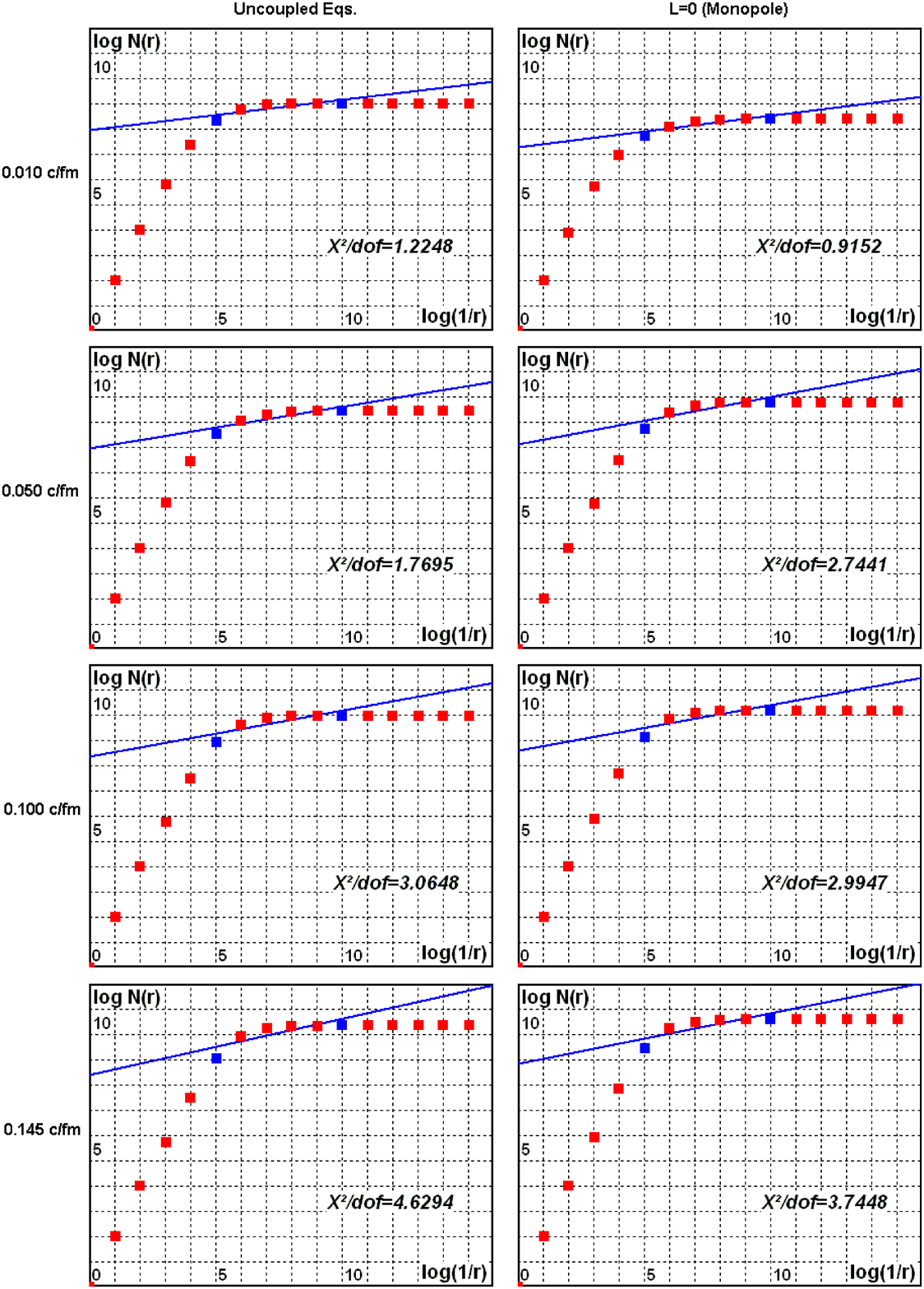} }
\caption{\label{fig:13}The fractal dimensions of the (radius $\leftrightarrow$ radial momentum) Poincare one-nucleon maps as linear slopes.}
\end{figure}

We calculated $d_{f}$ as slopes of the linear fits of $\log_{2}^{N\left(r\right)}$ \textit{versus} 
$n=\log_{2}^{1/r}$ points, where we took the inferior limit of $log (1/r)$ to be $5$, and the superior 
threshold according to the highest resolutions for which a saturation plateau appears: $N \neq N(r)$ 
(Fig. \ref{fig:13}). At resolutions above threshold, no additional information on the points distribution 
can be gained.

The points from the Poincare maps are distributed in such a way that up to a certain resolution ($32 %
\times 32$ pixels) the algorithm does not take into account the individual pixels. In this first region 
analyzed the alignment is quite good, denoting a correlation between points. The information is not 
uniformly distributed in the phase space but represents a "cloud" of points with $d_{f}$ greater than 
$1$. This large-scale region thus offers a global image of the Poincare maps.

Above $32 \times 32$ resolution appear the effects linked with the contribution of the individual dots 
of the Poincare maps, reflected by significant variations of the fractal dimensions. By choosing the 
points for the fitting procedure in such a way, we are also in agreement with the following criterion: 
the fitted points should be singled out so their associated resolutions are to be pertained to at least 
three successive decimal logarithmic intervals \cite{mandelbrot-82}. In Figure \ref{fig:13} we selected 
two points with resolutions belonging to the first interval $\left[10\div10^{2}\right]$, three points 
to $\left[10^{2}\div10^{3}\right]$, and we took one point in the $\left[10^{3}\div10^{4}\right]$ range, 
just until constant values $N$ are reached.

The small-scale $d_{f}$ thus computed (Table \ref{tab:1}) represent the filling degrees of a fine 
detailed phase space and can be correlated with the Shannon entropies calculated in \cite{felea-09b}.

\begin{table}
\caption{\label{tab:1}The $d_{f}$ of the $\left( r\leftrightarrow p_{r}\right)$ Poincare single-particle maps}
\begin{ruledtabular}
\begin{tabular}{ccc}
\hline\noalign{\smallskip}
Oscillation frequency & Uncoupled eqs. & Monopole case \\
\noalign{\smallskip}\hline\noalign{\smallskip}
$\Omega \:_{ad} = 0.010\: c/\rm{fm}$ & $0.1194$ & $0.1262$ \\
$\Omega \:_{ad} = 0.050\: c/\rm{fm}$ & $0.1666$ & $0.1870$ \\
$\Omega \:\;\ \   = 0.100\: c/\rm{fm}$ & $0.1840$ & $0.1790$ \\
$\Omega _{res} = 0.145\: c/\rm{fm}$ & $0.2227$ & $0.2010$ \\
\noalign{\smallskip}\hline
\end{tabular}
\end{ruledtabular}
\end{table}

It can be remarked that the fractal dimensions generally increase with the specific vibration frequencies, 
from the quasi-stationary equilibrium regime at adiabatic oscillations to the unstable chaotical states 
of the nucleonic system in the resonance domain. At the same time, it should be mentioned that for the 
monopole case the fractal dimension has an intermittent deportment close to the resonance phase of 
interaction, at $\Omega=0.1\ c/\rm{fm}$.

\subsection{Autocorrelation functions}

The transition toward a quasi-periodical, aperiodical or chaotic behaviour can be also analyzed with 
the autocorrelation function of the variables characteristic for the temporal evolution of the studied 
physical system. When the "nuclear billiard" evolves to chaos, essential changes can be noticed in 
the shape of the autocorrelation function of a specific variable (continuous or discrete distributed). 
This type of function measures the correlation between a sequence of signals, and is usually defined as:

\begin{equation}
C\left( \tau \right) = \lim_{T\rightarrow \infty} \frac{1}{T} \int^{T}_{0} \xi \left( t\right) \cdot \xi \left( t+\tau \right) \ dt;
\end{equation}

\begin{equation}
\xi \left( t\right) =x\left(t\right) - \lim_{T\rightarrow \infty} \frac{1}{T} \int^{T}_{0} x\left( t\right) \ dt;
\end{equation}

\begin{equation}
C\left( \tau \right) = \lim_{N\rightarrow \infty} \frac{1}{N} \sum^{N}_{n=0} \xi \left( t_{n}\right) \cdot \xi \left( t_{n+1}\right);
\end{equation}

\begin{equation}
\xi \left( t_{n}\right)
=x\left( t_{n}\right) - \lim_{N\rightarrow \infty} \frac{1}{N} \sum^{N}_{n=0} \xi \left( t_{n}\right).
\end{equation}

For a laminar regular regime this function either has a constant value or presents decreasing oscillations 
in time. During a chaotic phase, it falls in, having an exponentially decreasing behaviour for uncorrelated 
signals, as shown in \cite{roux-81}.

\begin{figure}
\resizebox{1.\hsize}{!}{\includegraphics{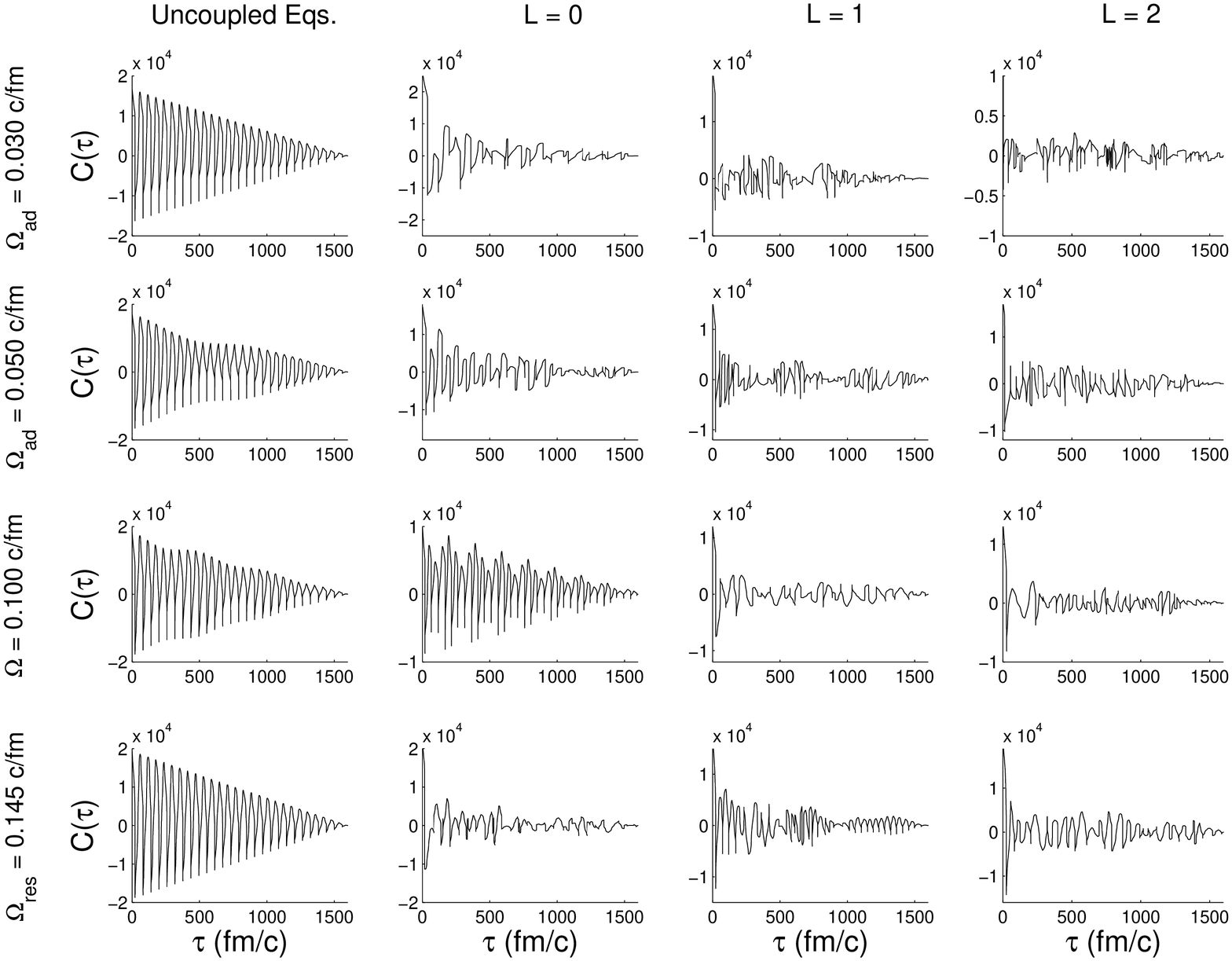} }
\caption{\label{fig:14}The autocorrelation function which correlates the radial momentum of a single-particle in time for various stages of nuclear interactions (from adiabatic to resonance regimes) and for several nuclear multipole deformations.}
\end{figure}

\begin{figure}
\resizebox{1.\hsize}{!}{\includegraphics{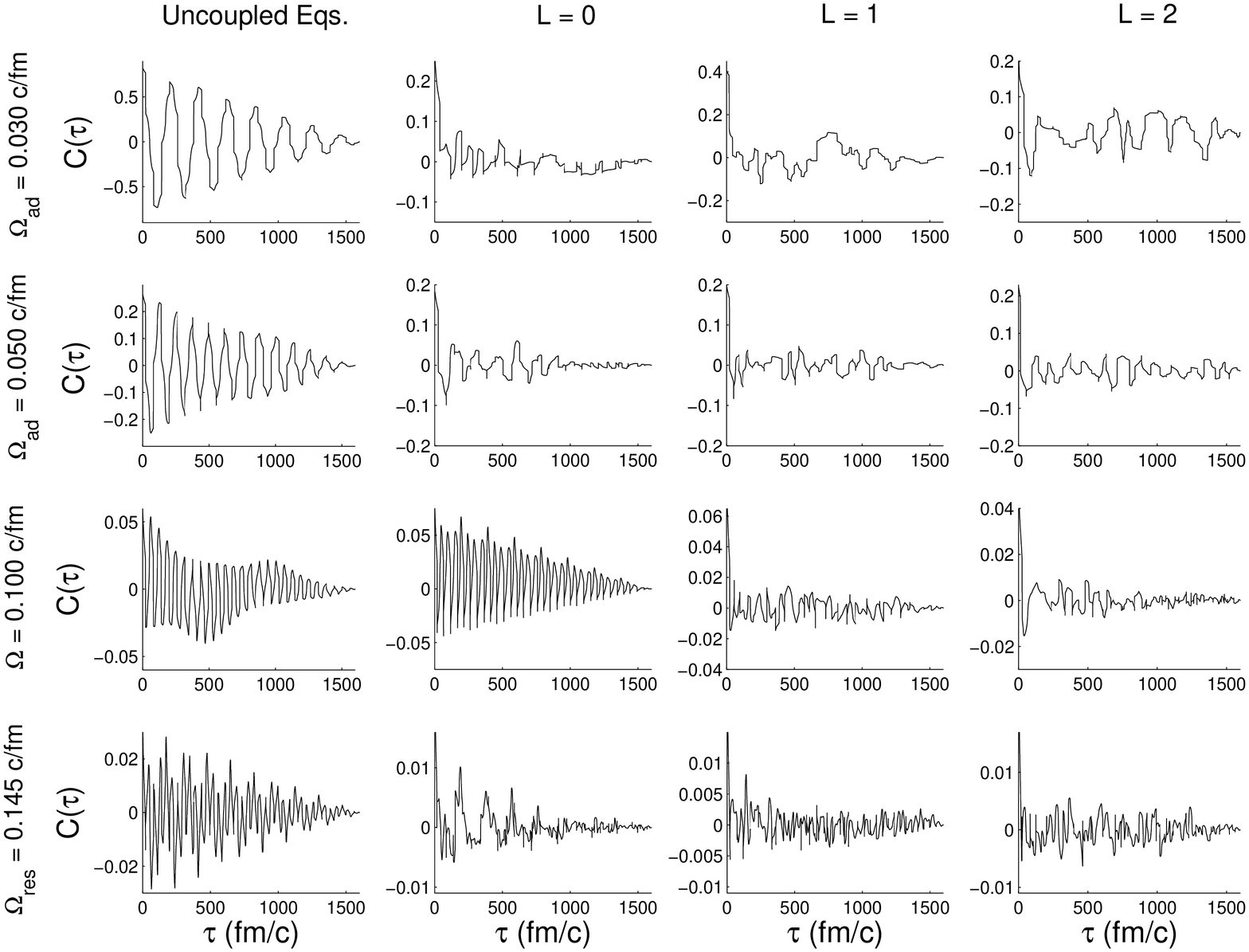} }
\caption{\label{fig:15}The autocorrelation function which correlates the collective variable in time for various stages of nuclear interactions (from adiabatic to resonance regimes) and for several nuclear multipole deformations.}
\end{figure}

We represented the autocorrelation function for the study of the single-particle (Fig. \ref{fig:14}) and 
collective dynamics (Fig. \ref{fig:15}). The analysis was thus done autocorrelating the radial momentum 
$p_{r}$ of the particle and respectively the collective variable $\alpha $ for the $UCE$ case and for 
all multipolarities, from all stages of interaction taken into account. One conclusion, that can be easily 
drawn from these figures, confirmed the previous results obtained with the other two methods. Thus, it 
can be noticed the increase of the chaotic bearing once the oscillation frequency of the potential well 
is varied from the adiabatic to the resonance regime, for all degrees of multipole.

When studying the case with single-particle degrees of freedom uncoupled from the collective ones, the 
dominant behaviour is aperiodical oscillation in time characteristic for a steady adiabatic phase of 
the interaction, even though the vibration frequency is varied. Only at resonance the shape is somewhat 
changed, indicating a greater degree of chaos.

As for the monopole to quadrupole deformations the decreasing of the autocorrelation function is indeed 
of the exponential type, steeper as the radian frequency of oscillation is raised to $\Omega _{res}=0.145\ c/\rm{fm}$.

The exception is again found in the monopole case at $\Omega =0.1\ c/\rm{fm}$ frequency of oscillation, 
when the intermittent phase of the interaction, reflected by aperiodical oscillations, points out a steady 
behaviour prior to the resonance regime.

\section{\label{sec:3}Conclusions}
A comparative study was done between the interesting physical regimes of nuclear interaction: adiabatic 
and resonance, giving at this level only a qualitative picture of the possible scenarios towards a 
pure deterministic behaviour of chaotic type of the studied nucleonic system.

We envisaged the single-nucleon dynamics in a Woods-Saxon potential. The coupling between individual and 
collective degrees of freedom was shown to generate different paths to chaos, according to the order of 
multipolarity. In general, for all vibrational modes at the resonance frequency of oscillation, the onset 
of chaotical behaviour was found to be earlier than at any other adiabatic vibrations of the 2D potential well.

Also, a phase alternation of periodical and chaotic dynamics was found in the monopole case of nuclear 
wall oscillation at $\Omega =0.1\ c/\rm{fm}$, revealing a laminar dynamics prior to the resonance stage 
of interaction.

In order to verify the aforementioned results, the study was completed with the inclusion of several 
quantitative analyses, inter alia we mention: power spectra, Shannon informational entropies, and Lyapunov 
exponents \cite{felea-09b}.

\begin{acknowledgments}
We are grateful to R.I. Nanciu, I.S. Zgur\u{a}, A.\c{S}. C\^{a}rstea, S. Zaharia, A. Ghea\c{t}\u {a}, 
A. Mitru\c{t}, M. Rujoiu, R. M\u{a}rginean, and I. Ion for stimulating discussions on this paper.
\end{acknowledgments}


\begin{thebibliography}{}
\bibitem{burgio-95} G.F. Burgio, M. Baldo, A. Rapisarda, and P. Schuck, Phys. Rev. C \textbf{52}, 2475 (1995).

\bibitem{baldo-96} M. Baldo, G.F. Burgio, A. Rapisarda, and P. Schuck, in \textit{Proceedings of the $XXXIV$ International Winter Meeting on Nuclear Physics, Bormio, Italy, 1996}, edited by I. Iori. arXiv:nucl-th/9602030

\bibitem{baldo-98} M. Baldo, G.F. Burgio, A. Rapisarda, and P. Schuck, Phys. Rev. C \textbf{58}, 2821 (1998).

\bibitem{blocki-78} J. Blocki, Y. Boneh, J.R. Nix, J. Randrup, M. Robel, A.J. Sierk, and W.J. Swiatecki, Ann. Phys. (N.Y.) \textbf{113}, 330 (1978).

\bibitem{ring-80} P. Ring and P. Schuck, \textit{The Nuclear Many Body Problem} (Springer-Verlag, Berlin, 1980) p. 388.

\bibitem{speth-81} J. Speth and A. van der Woude, Rep. Prog. Phys. \textbf{44}, 719 (1981).

\bibitem{wong-82} C.Y. Wong, Phys. Rev. C \textbf{25}, 1460 (1982).

\bibitem{grassberger-83} P. Grassberger and I. Procaccia, Phys. Rev. Lett. \textbf{50}, 346 (1983).

\bibitem{sieber-89} M. Sieber and F. Steiner, Physica D \textbf{44}, 248 (1990).

\bibitem{rapisarda-91} A. Rapisarda and M. Baldo, Phys. Rev. Lett. \textbf{66}, 2581 (1991).

\bibitem{abul-magd-91} A.Y. Abul-Magd and H.A. Weidenm\"{u}ller, Phys. Lett. B \textbf{261}, 207 (1991).

\bibitem{blocki-92} J. Blocki, F. Brut, T. Srokowski, and W.J. Swiatecki, Nucl. Phys. A\textbf{545}, 511c (1992).

\bibitem{blumel-92} R. Bl\"{u}mel and J. Mehl, J. Stat. Phys. \textbf{68}, 311 (1992).

\bibitem{baldo-93} M. Baldo, E.G. Lanza, and A. Rapisarda, Chaos \textbf{3}, 691 (1993).

\bibitem{blocki-93} J. Blocki, J.J. Shi, and W.J. Swiatecki, Nucl. Phys. A\textbf{554}, 387 (1993).

\bibitem{berry-93} M.V. Berry and J.M. Robbins, Proc. R. Soc., London, Sect. A \textbf{442}, 641 (1993).

\bibitem{ott-93} E. Ott, \textit{Chaos in Dynamical Systems} (Cambridge University Press, Cambridge, England, 1993).

\bibitem{bauer-94} W. Bauer, D. McGrew, V. Zelevinsky, and P. Schuck, Phys. Rev. Lett. \textbf{72}, 3771 (1994).

\bibitem{hilborn-94} R. Hilborn, \textit{Chaos and Nonlinear Dynamics} (Oxford University Press, Oxford, England, 1994).

\bibitem{blumel-94} R. Bl\"{u}mel and B. Esser, Phys. Rev. Lett. \textbf{72}, 3658 (1994).

\bibitem{drozdz-94} S. Drozdz, S. Nishizaki, and J. Wambach, Phys. Rev. Lett. \textbf{72}, 2839 (1994).

\bibitem{drozdz-95} S. Drozdz, S. Nishizaki, J. Wambach, and J. Speth, Phys. Rev. Lett. \textbf{74}, 1075 (1995).

\bibitem{bauer1-95} W. Bauer, D. McGrew, V. Zelevinsky, and P. Schuck, Nucl. Phys. A\textbf{583}, 93 (1995).

\bibitem{jarzynski-95} C. Jarzynski, Phys. Rev. Lett. \textbf{74}, 2937 (1995).

\bibitem{bulgac-95} A. Bulgac and D. Kusnezov, Chaos, Solitons and Fractals \textbf{5}, 1051 (1995).

\bibitem{atalmi-96a} A. Atalmi, M. Baldo, G.F. Burgio, and A. Rapisarda, Phys. Rev. C \textbf{53}, 2556 (1996). arXiv:nucl-th/9509020

\bibitem{atalmi-96b} A. Atalmi, M. Baldo, G.F. Burgio, and A. Rapisarda, in \textit{Proceedings of the $XXXIV$ International Winter Meeting on Nuclear Physics, Bormio, Italy, 1996}, edited by I. Iori. arXiv:nucl-th/9602039

\bibitem{papachristou-08} P.K. Papachristou, E. Mavrommatis, V. Constantoudis, F.K. Diakonos, and J. Wambach, Phys. Rev. C \textbf{77}, 044305 (2008). arXiv:nucl-th/0803.3336

\bibitem{felea-01} D. Felea, C. Be\c{s}liu, R.I. Nanciu, Al. Jipa, I.S. Zgur\u{a}, R. M\u{a}rginean, M. Haiduc, A. Ghea\c{t}\u{a}, and 
M. Ghea\c{t}\u{a}, in \textit{Proceedings of the $7^{th}$ International Conference "Nucleus-Nucleus Collisions", Strasbourg, 2000}, edited by 
W. Norenberg \textit{et al.} (North-Holland, Amsterdam, The Netherlands, 2001) p. 222.

\bibitem{felea-02} D. Felea, \textit{The Study of Nuclear Fragmentation Process in Nucleus-Nucleus Collisions at Energies higher than 1 A GeV}, Ph.D. thesis, University of Bucharest, Faculty of Physics (2002) p. 134.

\bibitem{bordeianu-08a} C.C. Bordeianu, C. Be\c{s}liu, Al. Jipa, D. Felea, and I.V. Grossu, Comput. Phys. Commun. \textbf{178}, 788 (2008).

\bibitem{bordeianu-08b} C.C. Bordeianu, D. Felea, C. Be\c{s}liu, Al. Jipa, and I.V. Grossu, Comput. Phys. Commun. \textbf{179}, 199 (2008).

\bibitem{bordeianu-08c} C.C. Bordeianu, D. Felea, C. Be\c{s}liu, Al. Jipa, and I.V. Grossu, Rom. Rep. in Phys. \textbf{60}, 287 (2008).

\bibitem{pomeau-80} Y. Pomeau and P. Manneville, Commun. Math. Phys. \textbf{74}, 189 (1980).

\bibitem{berge-80} P. Berge, M. Dubois, P. Manneville, and Y. Pomeau, J. Phys. (Paris) \textbf{41}, L344 (1980).

\bibitem{pomeau-81} Y. Pomeau, J.C. Roux, A. Rossi, S. Bachelart, and C. Vidal, J. Phys. (Paris) \textbf{42}, L271 (1981).

\bibitem{linsay-81} P.S. Linsay, Phys. Rev. Lett. \textbf{47}, 1349 (1981).

\bibitem{testa-82} J. Testa, J. Perez, and C. Jeffries, Phys. Rev. Lett. \textbf{48}, 714 (1982).

\bibitem{jeffries-82} C. Jeffries and J. Perez, Phys. Rev. A \textbf{26}, 2117 (1982).

\bibitem{dubois-83} M. Dubois, M.A. Rubio, and P. Berge, Phys. Rev. Lett. \textbf{51}, 1446 (1983).

\bibitem{yeh-83} W.J. Yeh and Y.H. Kao, Appl. Phys. Lett. \textbf{42}, 299 (1983).

\bibitem{huang-87} J.Y. Huang and J.J. Kim, Phys. Rev. A \textbf{36}, 1495 (1987).

\bibitem{richetti-87} P. Richetti, P. DeKepper, J.C. Roux, and H.L. Swinney, J. Stat. Phys. \textbf{48}, 977 (1987).

\bibitem{kreisberg-91} N. Kreisberg, W.D. McCormick, and H.L. Swinney, Physica D \textbf{50}, 463 (1991).

\bibitem{dahlqvist-92} P. Dahlqvist, J. Phys. A \textbf{25}, 6265 (1992).

\bibitem{dahlqvist-95} P. Dahlqvist, Nonlinearity \textbf{8}, 11 (1995).

\bibitem{artuso-96} R. Artuso, G. Casati, and I. Guarneri, J. Stat. Phys. \textbf{83}, 977 (1996).

\bibitem{altmann-07} E.G. Altmann, \textit{Intermittent Chaos in Hamiltonian Dynamical Systems}, Ph.D. thesis, Max Planck Institute for the Physics of Complex Systems in Dresden (2007), WUB-DIS 2007-02.

\bibitem{bunimovich-01} L.A. Bunimovich, Chaos \textbf{11}, 802 (2001).

\bibitem{bunimovich-03} L.A. Bunimovich, Chaos \textbf{13}, 903 (2003).

\bibitem{altmann-05} E.G. Altmann, A.E. Motter, and H. Kantz, Chaos \textbf{15}, 033105 (2005). arXiv:nlin/0502058v2

\bibitem{altmann-06} E.G. Altmann, A.E. Motter, and H. Kantz, Phys. Rev. E \textbf{73}, 026207 (2006). arXiv:nlin/0601008v1

\bibitem{porter-06} M.A. Porter and S. Lansel, Notices of the AMS \textbf{53}, 334 (2006).

\bibitem{saito-82} N. Sait$\hat{o}$, H. Hirooka, J. Ford, F. Vivaldi, and G.H. Walker, Physica D \textbf{5}, 273 (1982).

\bibitem{markus-03} M. Markus and M. Schmick, Physica A \textbf{328}, 335 (2003).

\bibitem{bauer2-95} W. Bauer, Phys. Rev. C \textbf{51}, 803 (1995).

\bibitem{felea-99} C. Be\c{s}liu, D. Felea, V. Topor-Pop, A. Ghea\c{t}\u{a}, I.S. Zgur\u{a}, Al. Jipa, and R. Zaharia, Phys. Rev. C \textbf{60}, 
024609 (1999).

\bibitem{pochodzalla-87} J. Pochodzalla \textit{et al.}, Phys. Rev. C \textbf{35}, 1695 (1987).

\bibitem{kunde-91} G.J. Kunde \textit{et al.}, Phys. Lett. B \textbf{272}, 202 (1991).

\bibitem{morrissey-94} D.J. Morrissey, W. Benenson, and W.A. Friedman, Annu. Rev. Nucl. Part. Sci. \textbf{44}, 27 (1994).

\bibitem{serfling-98} V. Serfling \textit{et al.}, Phys. Rev. Lett. \textbf{80}, 3928 (1998).

\bibitem{schuster-84} H.G. Schuster, \textit{Deterministic Chaos: an introduction} (Physik-Verlag, Weinheim, Federal Republic of Germany, 1984).

\bibitem{poincare-890} H.J. Poincare, Acta Math. \textbf{13}, (1890).

\bibitem{poincare-892} H.J. Poincare, \textit{Les Methodes Nouvelles de la Mechanique Celeste} (Gauthier-Villars, Paris, 1892); 
English version: \textit{N.A.S.A. Translation TT F-450/452} (U.S. Fed. Clearinghouse, Springfield, VA, U.S.A., 1967).

\bibitem{grossu-09a} I.V. Grossu, C. Be\c{s}liu, M.V. Rusu, Al. Jipa, C.C. Bordeianu, and D. Felea, Comput. Phys. Commun. \textbf{180}, 1999 (2009). arXiv:0901.4643

\bibitem{grossu-09b} I.V. Grossu, D. Felea, C. Be\c{s}liu, Al. Jipa, C.C. Bordeianu, E. Stan, and T. E\c{s}anu, Comput. Phys. Commun., DOI: 10.1016/j.cpc.2009.12.005 (in press).

\bibitem{landau-07} R.H. Landau, M.J. P$\acute{a}$ez, and C.C. Bordeianu, \textit{Computational Physics - Problem Solving with Computers} (WILEY-VCH Verlag GmbH Et Co. KGaA, Weinheim, Germany, 2007) p. 304.

\bibitem{landau-08} R.H. Landau, M.J. P$\acute{a}$ez, and C.C. Bordeianu, \textit{A Survey of Computational Physics - Introductory Computational Science} (Princeton University Press, Princeton, New Jersey, USA, and Woodstock, Oxfordshire, UK, 2008) p. 335.

\bibitem{mandelbrot-82} B.B. Mandelbrot, \textit{The Fractal Geometry of Nature} (W.H. Freeman and Co., San Francisco, CA, U.S.A., 1982). Revised edition of \textit{Fractals: Form, Chance and Dimension} (W.H. Freeman and Co., San Francisco, CA, U.S.A., 1977).

\bibitem{felea-09b} D. Felea, C.C. Bordeianu, I.V. Grossu, C. Be\c{s}liu, Al. Jipa, A.A. Radu, and E. Stan, "Intermittency route to chaos for the nuclear billiard - a quantitative study", Phys. Rev. C (submitted).

\bibitem{roux-81} J.C. Roux, A. Rossi, S. Bachelart, and C. Vidal, Physica D \textbf{2}, 395 (1981).
\end{thebibliography}
\end{document}